\documentclass[journal,oneside, onecolumn, draftclsnofoot]{IEEEtran}
\usepackage{cite}

\usepackage{epstopdf}					
\ifCLASSINFOpdf
  \usepackage[pdftex]{graphicx}
  \graphicspath{{figures/}}
\else
  \usepackage[dvips]{graphicx}
  \graphicspath{{figures/}}
\fi
\usepackage[cmex10]{amsmath}
\usepackage{amssymb}
\usepackage{algorithm}
\usepackage{algpseudocode}

\usepackage{mdwmath}
\usepackage{mdwtab}

\usepackage[belowskip=0pt,aboveskip=0pt,small]{caption}
\IEEEaftertitletext{\vspace{-2.5\baselineskip}}
\begin{document}

%
\title{Joint Optimization of Bit and Power Loading for Multicarrier Systems}
%
%
%


\author{Ebrahim Bedeer,
        Octavia A. Dobre, 
        Mohamed~H.~Ahmed,
       and~Kareem~E.~Baddour
}

\IEEEpubid{0000--0000/00\$00.00~\copyright~2012 IEEE}


\maketitle

\begin{abstract}
In this letter, a novel low complexity bit and power loading algorithm is formulated for multicarrier communication systems. The proposed algorithm jointly maximizes the throughput and minimizes the transmit power through a weighting coefficient $\alpha$, while meeting constraints on the target bit error rate (BER) per subcarrier and on the total transmit power. The optimization problem is solved by the Lagrangian multiplier method if the initial $\alpha$ causes the transmit power not to violate the power constraint; otherwise, a bisection search is used to find the appropriate $\alpha$. Closed-form expressions are derived for the close-to-optimal bit and power allocations per subcarrier, average throughput, and average transmit power.  Simulation results illustrate the performance of the proposed algorithm and demonstrate its superiority with respect to existing allocation algorithms. Furthermore, the results show that the performance of the proposed algorithm approaches that of the exhaustive search for the discrete optimal allocations.
\end{abstract}


%

\vspace*{-10pt}
\section{Introduction}
\vspace{-1pt}
\IEEEPARstart{M}{ulticarrier} modulation is recognized as a robust and efficient transmission technique, as evidenced by its consideration for diverse communication systems and adoption by several wireless standards \cite{fazel2008multi}. The performance of multicarrier communication systems can be significantly improved by dynamically adapting the transmission parameters, such as power, constellation size, symbol rate, coding rate/scheme, or any combination of these, according to the channel quality or the wireless standard specifications \cite{wyglinski2005bit,  liu2009adaptive, mahmood2010efficient, willink1997optimization}. 

To date, most of the research literature has focused on the single objective of either maximizing the throughput or minimizing the transmit power separately (see, e.g., \cite{wyglinski2005bit,  liu2009adaptive, willink1997optimization, mahmood2010efficient} and references therein). In \cite{wyglinski2005bit}, Wyglinski \textit{et al.}  proposed an incremental bit loading algorithm with uniform power in order to maximize the throughput while guaranteeing a target BER. Liu and Tang \cite{liu2009adaptive} proposed a power loading algorithm with uniform bit loading that aims to minimize the transmit power while guaranteeing a target~BER. In \cite{mahmood2010efficient}, Mahmood and Belfiore proposed an efficient greedy bit allocation algorithm that minimizes the transmit power subject to fixed throughput and BER per subcarrier constraints.

In emerging wireless communication systems, various requirements are needed. For example, maximizing the throughput is favoured if sufficient guard bands exist to separate users, while minimizing the transmit power is prioritized when operating in interference-prone shared spectrum environments, to prolong the battery life time of battery-operated nodes, as well as to support environmentally-friendly transmission behaviors.
This motivates us to formulate a multiobjective optimization (MOOP) problem that optimizes the conflicting and incommensurable throughput and power objectives. According to the MOOP principle, there is no solution that improves one of the objectives without deteriorating others. Therefore, MOOP produces a set of optimal solutions and it is the responsibility of the resource allocation entity to choose the most preferred optimal solution depending on its preference \cite{miettinen1999nonlinear}. A well known approach to solve MOOP problems is to linearly combine the competing objective functions into a single objective function, through weighting coefficients that reflect the required preferences \cite{miettinen1999nonlinear}. These preferences can be prescribed and fixed during the solution process (as in posteriori and priori methods) or can be changed during the solution process (interactive methods) \cite{miettinen1999nonlinear}. In this paper, we adopt an interactive approach in order to obtain a low complexity solution.

We propose a low complexity algorithm that jointly maximizes the throughput and minimizes the total transmit power, subject to constraints on the BER per subcarrier and the total transmit power. Limiting the total transmit power is crucial for a variety of reasons, e.g.,  to reflect the transmitter's power amplifier limitations, to satisfy regulatory maximum power limits, and to limit interference/ encourage frequency reuse. Moreover, including the total subcarrier power in the objective function is especially desirable, as it minimizes the transmit power when the power constraint is inactive.
 Closed-form expressions are derived for the close-to-optimal bit and power allocations, average throughput, and average transmit power. Simulation results show that the proposed algorithm outperforms existing bit and power loading schemes in the literature, while requiring similar or reduced computational effort. The results also indicate that the proposed algorithm's performance approaches that of the exhaustive search for the optimal discrete allocations, with significantly reduced computational effort.



\IEEEpubidadjcol

\section{Proposed Link Adaptation Scheme} \label{sec:opt}
\subsection{Optimization Problem Formulation}
A multicarrier communication system decomposes the signal bandwidth into a set of $N$ orthogonal narrowband subcarriers of equal bandwidth. Each subcarrier $i$ transmits $b_i$ bits using power $\mathcal{P}_i$, $i = 1, ..., N$. Following the common practice in the literature, a delay- and error-free feedback channel is assumed to exist between the transmitter and receiver for reporting the channel state information \cite{liu2009adaptive, willink1997optimization, mahmood2010efficient}.

In order to maximize the throughput and minimize the transmit power  subject to BER and total transmit power constraints, the optimization problem is  formulated as
\begin{IEEEeqnarray}{c}
\underset{b_i}{\textup{Maximize}} \quad \sum_{i = 1}^{N}b_i \quad \textup{and} \quad \underset{\mathcal{P}_i}{\textup{Minimize}} \quad \sum_{i = 1}^{N}\mathcal{P}_i, \nonumber
\end{IEEEeqnarray}
\begin{IEEEeqnarray}{RCL}
\textup{subject to} &{} \quad {}& \textup{BER}_i \leq \textup{BER}_{th,i}, \nonumber \\
 &{} \quad {}& \sum_{i = 1}^{N} \mathcal{P}_i \leq \mathcal{P}_{th}, \qquad i = 1, ..., N, \label{eq:eq_first}
\end{IEEEeqnarray}
where $\textup{BER}_i$ and $\textup{BER}_{th,i}$ are the BER and threshold value of BER per subcarrier\footnote{The constraint on the BER per subcarrier is a suitable formulation that results in similar BER characteristics compared to an average BER constraint, especially at high signal-to-noise ratios (SNRs)~\cite{willink1997optimization}. Further, it significantly reduces the computational complexity by yielding closed-form expressions.} $i$, $i$ = 1, ..., $N$, respectively, and $\mathcal{P}_{th}$ is the total transmit power threshold. An approximate expression for the BER per subcarrier $i$ for $M$-ary QAM is given by \cite{liu2009adaptive}
\begin{IEEEeqnarray}{rCl}
\textup{BER}_i &{} \approx  {}& 0.2 \: \textup{exp}\left (-1.6 \: \frac{\mathcal{P}_i}{2^{b_i} - 1} \frac{\left | \mathcal{H}_i \right |^2}{\sigma^2_n} \right ), \label{eq:BER}
\end{IEEEeqnarray}
where $ \mathcal{H}_i $ is the channel gain of subcarrier $i$ and $\sigma^2_n$ is the variance of the additive white Gaussian noise (AWGN).
The multi-objective optimization function in (\ref{eq:eq_first}) can be rewritten as a linear combination of multiple objective functions as follows
\setlength\abovedisplayskip{0pt}
\begin{IEEEeqnarray}{c}
\underset{\mathcal{P}_i , b_i}{\textup{Minimize}}  \quad  \mathcal{F}(\mathbf{p},\mathbf{b})  = \: \alpha \sum_{i = 1}^{N}\mathcal{P}_i - (1-\alpha)\sum_{i = 1}^{N}b_i, \label{eq:p1} \nonumber \\
\textup{subject to} \quad  g_{\varrho}(\mathcal{P}_i,b_i)  \hfill \nonumber \\
= \left\{\begin{matrix}
 0.2 \: \textup{exp}\left ( \frac{- 1.6 \: \mathcal{C}_i \mathcal{P}_i}{2^{b_i} - 1} \right ) - \textup{BER}_{th,i} \leq 0, \quad  \varrho = 1, ..., N,  \\
\sum_{i = 1}^{N} \mathcal{P}_i \leq \mathcal{P}_{th}, \hfill \varrho = N+1,
\end{matrix}\right.
 \label{eq:ineq_const}
\end{IEEEeqnarray}
where $\alpha$ ($0 < \alpha < 1$) is a weighting coefficient which indicates the rate at which the multicarrier system is willing to trade off the values of the objective functions in order to obtain a low complexity solution \cite{miettinen1999nonlinear} (i.e., a higher value of $\alpha$ favors minimizing the transmit power, whereas a lower value of $\alpha$ favors maximizing the throughput).
$\mathcal{C}_i = \: \frac{\left | \mathcal{H}_i \right |^2}{\sigma^2_n}$ is the channel-to-noise ratio for subcarrier $i$, and $\mathbf{p} = [\mathcal{P}_1, ..., \mathcal{P}_N]^T$ and $\mathbf{b} = [b_1, ..., b_N]^T$ are the \textit{N}-dimensional power and bit distribution vectors, respectively, with $[.]^T$ denoting the transpose operation.

\vspace{-17pt}

\subsection{Bit and Power Allocations}
\vspace{-5pt}
The optimization problem in (\ref{eq:ineq_const}) can be solved numerically
; however, this is computationally complex. A low complexity solution can be obtained by relaxing the power constraint in (\ref{eq:ineq_const}), i.e., $\varrho \neq N+1$, and then applying the method of Lagrange multipliers. Accordingly,  the inequality constraints are transformed to equality constraints by adding non-negative slack variables, $\mathcal{Y}_{i}^2$, $ \varrho = i = 1, ..., N$ \cite{rao2009engineering}. Hence, the constraints are given as
\setlength{\arraycolsep}{0.0em}
\begin{IEEEeqnarray}{RCL}
\mathcal{G}_{i}(\mathbf{p},\mathbf{b},\mathbf{y}) &{} = {}& g_{i}(\mathbf{p},\mathbf{b}) + \mathcal{Y}_{i}^2 = 0, \quad i = 1, ..., N,
\label{eq:eq_const}
\end{IEEEeqnarray}
where $\mathbf{y} = [\mathcal{Y}_1^2, ..., \mathcal{Y}_N^2]^T$ is the vector of slack variables, and the Lagrange function $\mathcal{L}$ is expressed as
\vspace{-1pt}
\begin{IEEEeqnarray}{c}
\mathcal{L}(\mathbf{p},\mathbf{b},\mathbf{y},\boldsymbol\lambda)  =  \mathcal{F}(\mathbf{p},\mathbf{b}) + \sum_{i = 1}^{N} \lambda_{i} \: \mathcal{G}_{i}(\mathbf{p},\mathbf{b},\mathbf{y}), \nonumber
\end{IEEEeqnarray} \vspace{-5pt}
\begin{IEEEeqnarray}{RCL}
 & = &  \alpha \sum_{i = 1}^{N}\mathcal{P}_i - (1-\alpha)\sum_{i = 1}^{N}b_i \hspace*{0.7cm} \nonumber \\
 & + &  \sum_{i = 1}^{N} \lambda_{i}\:\Bigg[ 0.2 \: \textup{exp}\left ( \frac{- 1.6 \: \mathcal{C}_i \mathcal{P}_i}{2^{b_i} - 1} \right ) - \textup{BER}_{th,i} + \mathcal{Y}_{i}^2\Bigg],
\end{IEEEeqnarray}
where $\boldsymbol\lambda = [\lambda_1, ..., \lambda_N]^T$ is the vector of Lagrange multipliers. A stationary point is found when $\nabla \mathcal{L}(\mathbf{p},\mathbf{b},\mathbf{y},\boldsymbol\lambda) = 0$ ($\nabla$ denotes the gradient), which yields
\begin{IEEEeqnarray}{RCL}
\frac{\partial \mathcal{L}}{\partial \mathcal{P}_i} &{} = {}& \alpha - 0.2 \: \lambda_i \frac{1.6 \: \mathcal{C}_i}{2^{b_i}-1} \: \textup{exp}\left ( \frac{- 1.6 \: \mathcal{C}_i \mathcal{P}_i}{2^{b_i} - 1} \right ) = 0,\label{eq:eq1}
\end{IEEEeqnarray}
\begin{IEEEeqnarray}{RCL}
\frac{\partial \mathcal{L}}{\partial b_i} &{} = {}& -(1 - \alpha) + 0.2 \ln (2) \: \lambda_i \frac{1.6 \: \mathcal{C}_i \mathcal{P}_i 2^{b_i}}{(2^{b_i}-1)^2} \: \nonumber \\ & & \hfill \textup{exp}\left ( \frac{- 1.6 \: \mathcal{C}_i \mathcal{P}_i}{2^{b_i} - 1} \right ) = 0,\label{eq:eq2}\\
\frac{\partial \mathcal{L}}{\partial \lambda_i} & = & 0.2 \: \textup{exp}\left ( \frac{- 1.6 \: \mathcal{C}_i \mathcal{P}_i}{2^{b_i} - 1} \right ) - \textup{BER}_{th,i} + \mathcal{Y}_i^2= 0, \label{eq:eq3} \\
\frac{\partial \mathcal{L}}{\partial \mathcal{Y}_i} & = & 2\lambda_i \mathcal{Y}_i = 0. \label{eq:eq4}
\end{IEEEeqnarray}
It can be seen that (\ref{eq:eq1}) to (\ref{eq:eq4}) represent $4N$ equations in the $4N$ unknown components of the vectors $\mathbf{p}, \mathbf{b}, \mathbf{y}$, and $\boldsymbol\lambda$. By solving (\ref{eq:eq1}) to (\ref{eq:eq4}), one obtains the solution $\mathbf{p}^*, \mathbf{b}^*$. Equation (\ref{eq:eq4}) implies that either $\lambda_i$ = 0 or $\mathcal{Y}_i$ = 0; hence, two possible cases exist and we are going to investigate each case independently.

--- \textit{Case 1}: Setting $\lambda_i = 0$ in (\ref{eq:eq1}) to (\ref{eq:eq4}) results in an underdetermined system of $N$ equations in $3N$ unknowns, and, hence, no unique solution can be reached.

--- \textit{Case 2}: Setting $\mathcal{Y}_i = 0$ in (\ref{eq:eq1}) to (\ref{eq:eq4}), we can relate $\mathcal{P}_i$ and $b_i$ from (\ref{eq:eq1}) and (\ref{eq:eq2}) as follows
\setlength{\arraycolsep}{0.0em}
\begin{IEEEeqnarray}{RCL}
\mathcal{P}_i &{} = {}& \frac{1- \alpha}{\alpha \ln(2)}(1 - 2^{-b_i}), \label{eq:eq8}
\end{IEEEeqnarray}
with $\mathcal{P}_i \geq 0$ if and only if $b_i \geq 0$. By substituting (\ref{eq:eq8}) into (\ref{eq:eq3}), one obtains the solution
\setlength{\arraycolsep}{0.0em}
\begin{IEEEeqnarray}{c}
b_i^* = \frac{1}{\log(2)}\log\Bigg[- \frac{1-\alpha }{\alpha \ln(2)} \frac{1.6 \: \mathcal{C}_i}{\ln(5 \: \textup{BER}_{th,i})}\Bigg]. \label{eq:eq10}
\end{IEEEeqnarray}
Consequently, from (\ref{eq:eq8}) one gets
\setlength{\arraycolsep}{0.0em}
\begin{IEEEeqnarray}{c}
\mathcal{P}_i^* = \frac{1-\alpha }{\alpha \ln(2)}\Bigg[ 1 - \Big(- \frac{1-\alpha }{\alpha \ln(2)} \frac{1.6 \: \mathcal{C}_i}{\ln(5 \: \textup{BER}_{th,i})} \Big)^{-1} \Bigg]. \label{eq:eq12}
\end{IEEEeqnarray}
Since we consider $M$-ary QAM, $b_i$ should be greater than 2. From (\ref{eq:eq10}), to have $b_i \geq 2$, $\mathcal{C}_i$ must satisfy the condition
\setlength{\arraycolsep}{0.0em}
\begin{IEEEeqnarray}{c}
\mathcal{C}_i \geq \mathcal{C}_{th,i} = \frac{4}{1.6} \frac{\alpha \ln(2)}{1-\alpha} (-\ln(5\textup{BER}_{th,i})), i = 1, ..., N. \label{eq:condition} \IEEEeqnarraynumspace
\end{IEEEeqnarray}

The relaxed optimization problem is not convex and, hence, the Karush-Kuhn-Tucker (KKT) conditions do not guarantee that ($\mathbf{p}^*,\mathbf{b}^*$) represents a global optimum \cite{rao2009engineering}; the proof of the KKT conditions is not provided due to the space limitations. To characterize the gap to the global optimum solution, we compare the obtained local optimum results to the global optimum results obtained through the exhaustive search in the next section.


If the total transmit power $\sum_{i = 1}^{N} \mathcal{P}_i$ is below $\mathcal{P}_{th}$, then the final bit and power allocations are reached. On the other hand, if the transmit power exceeds $\mathcal{P}_{th}$, the algorithm adopts the interactive approach and overrides the initial value of $\alpha$ to meet the power constraint.  This is achieved by giving more weight to the transmit power minimization in (\ref{eq:ineq_const}), i.e., by increasing $\alpha$. The lowest $\alpha^*$ that satisfies the constraint, i.e., $\alpha^*$ that results in the highest total power which is lower than $\mathcal{P}_{th}$, is found through the bisection search\footnote{This is true as the total transmit power calculated from (\ref{eq:eq12}) is a decreasing function of $\alpha$. The proof is not provided due to the space limitations.} (please note that lower values of $\alpha$ produce lower values of the objective function in (\ref{eq:ineq_const})). The proposed algorithm can be formally stated as follows:

\vspace*{-10pt}
\floatname{algorithm}{}
\begin{algorithm}
\renewcommand{\thealgorithm}{}
\caption{\textbf{Proposed Algorithm}}
\begin{algorithmic}[1]
\small
\State \textbf{INPUT} The AWGN variance ($\sigma^2_n$), channel gain per subcarrier $i$ ($\mathcal{H}_i$), target BER per subcarrier $i$ ($\textup{BER}_{th,i}$), initial weighting parameter $\alpha$, and tolerance $\epsilon$.
\For{$i$ = 1, ..., $N$}
%
\If{$\mathcal{C}_i \geq \mathcal{C}_{th,i} =  - \frac{4}{1.6} \: \frac{\alpha \ln(2)}{1-\alpha} \: \ln(5\:\textup{BER}_{th,i})$}
\State  - $b_i^*$ and $\mathcal{P}_i^*$ are given by (\ref{eq:eq10}) and (\ref{eq:eq12}), respectively.
\State - $b^*_{i,final}$ $\leftarrow$ Round $b_i^*$ to the nearest integer.
\State - $\mathcal{P}^*_{i,final}$ $\leftarrow$ Recalculate $\mathcal{P}_i^*$ according to (\ref{eq:BER}).
\Else
\State Null the corresponding subcarrier $i$.
\EndIf
\EndFor
\While{$\sum_{i = 1}^{N}\mathcal{P}^*_{i,final} - \mathcal{P}_{th} > \epsilon$}
\State - Set $\alpha_L$ = $\alpha$ and $\alpha_U$ = 1.
\State - Set $\alpha^* = (\alpha_L + \alpha_U)/2$.
\State  - Repeat steps: 3 to 9.
\If{$\sum_{i = 1}^{N}\mathcal{P}^*_{i,final} < \mathcal{P}_{th}$}
\State - Set $\alpha_U$ = $\alpha^*$, then $\alpha^* = (\alpha_L + \alpha_U)/2$.
\State  - Repeat steps: 3 to 9.
\Else
\State - Set $\alpha_L$ = $\alpha^*$, then $\alpha^* = (\alpha_L + \alpha_U)/2$.
\State  - Repeat steps: 3 to 9.
\EndIf
\EndWhile
\State \textbf{OUTPUT} $b^*_{i,final}$ and $\mathcal{P}^*_{i,final}$, $i$ = 1, ..., $N$.
\end{algorithmic}
\end{algorithm}

\vspace*{-19pt}
\subsection{Analytical Expressions of Average Throughput and Transmit Power}
\vspace*{-3pt}
When the initial value of $\alpha$ results in an inactive power constraint, the closed-form expressions for the average throughput and transmit power can be found by averaging the bit and power allocations given by (\ref{eq:eq10}) and (\ref{eq:eq12}), respectively, over $\mathcal{C}_i$. In such a case, the average throughput is expressed as
\begin{IEEEeqnarray}{RCL}
\textup{Throughput}_\textup{av}&{} = {}& \sum_{i = 1}^{N} \mathbb{E}\{b_i(\mathcal{C}_i)\} \nonumber \\
 &{} = {}& \sum_{i = 1}^{N} \int_{\mathcal{C}_{th,i}}^{\infty} b_i(\mathcal{C}_i)\Big[\leftthreetimes \textup{exp}(-\leftthreetimes \mathcal{C}_i)\Big] d\mathcal{C}_i, \label{eq:av_b} \IEEEeqnarraynumspace
\end{IEEEeqnarray}
where $\leftthreetimes \textup{exp}(-\leftthreetimes \mathcal{C}_i)$ is the exponential distribution of~$\mathcal{C}_i$ with mean $\frac{1}{\leftthreetimes}$, given that the channel gain $\mathcal{H}_i$ has a Rayleigh distribution. The integration in (\ref{eq:av_b}) is solved by parts yielding
\begin{IEEEeqnarray}{c}
\textup{Throughput}_\textup{av} = \sum_{i = 1}^{N} \frac{1}{\textup{log}(2)} \Bigg[\textup{log}(4) \: \textup{exp}(-\leftthreetimes \mathcal{C}_{th,i})  \nonumber \\
\hfill - \frac{\textup{Ei}(-\leftthreetimes \mathcal{C}_{th,i})}{\textup{ln}(10)} \Bigg], \label{eq:av_th}
\end{IEEEeqnarray}
where $\textup{Ei}(-z) = - \int_{z}^{\infty}\frac{\textup{e}^{-t}}{t} \: dt, \: z > 0$ is the exponential integral function. Similarly, the average transmit power is given by
\vspace{-2pt}
\begin{IEEEeqnarray}{c}
\textup{Power}_\textup{av} = \sum_{i = 1}^{N} \frac{1-\alpha}{\alpha \: \textup{ln}(2)} \Bigg[ \textup{exp}(-\leftthreetimes \mathcal{C}_{th,i}) \hspace*{1cm} \nonumber \\
\hfill + \frac{\leftthreetimes \: \mathcal{C}_{th,i}}{4} \textup{Ei}(-\leftthreetimes \mathcal{C}_{th,i}) \Bigg]. \label{eq:av_power}
\end{IEEEeqnarray}



\section{Simulation Results} \label{sec:sim}
This section investigates the performance of the proposed algorithm, and compares its performance with bit and power loading algorithms presented in the literature, as well as with the exhaustive search for the discrete global optimal allocations. The computational complexity of the proposed algorithm is also compared to the other schemes.

\subsection{Simulation Setup}
As an example of a multicarrier system, we consider orthogonal frequency division multiplexing (OFDM) with $N$ = 128 subcarriers. Without loss of generality, the BER constraint per subcarrier, $\textup{BER}_{th,i}$, is assumed to be the same for all subcarriers and set to $10^{-4}$. A Rayleigh fading environment with average channel power gain $\mathbb{E}\{\left | \mathcal{H}_i \right |^2\}$ = 1 is considered.
Representative results are presented, which were obtained through Monte Carlo trials for $10^{4}$ channel realizations with
$\epsilon = 10^{-9}$ mW and initial $\alpha = 0.5$. The transmit power objective function is scaled during simulations so that it is approximately within the same range as the throughput \cite{miettinen1999nonlinear}. For convenience, presented numerical results are displayed in the original scales.

\vspace*{-9pt}
\subsection{Performance of the Proposed Algorithm}
\begin{figure}[t]
	\centering
		\includegraphics[width=0.48\textwidth]{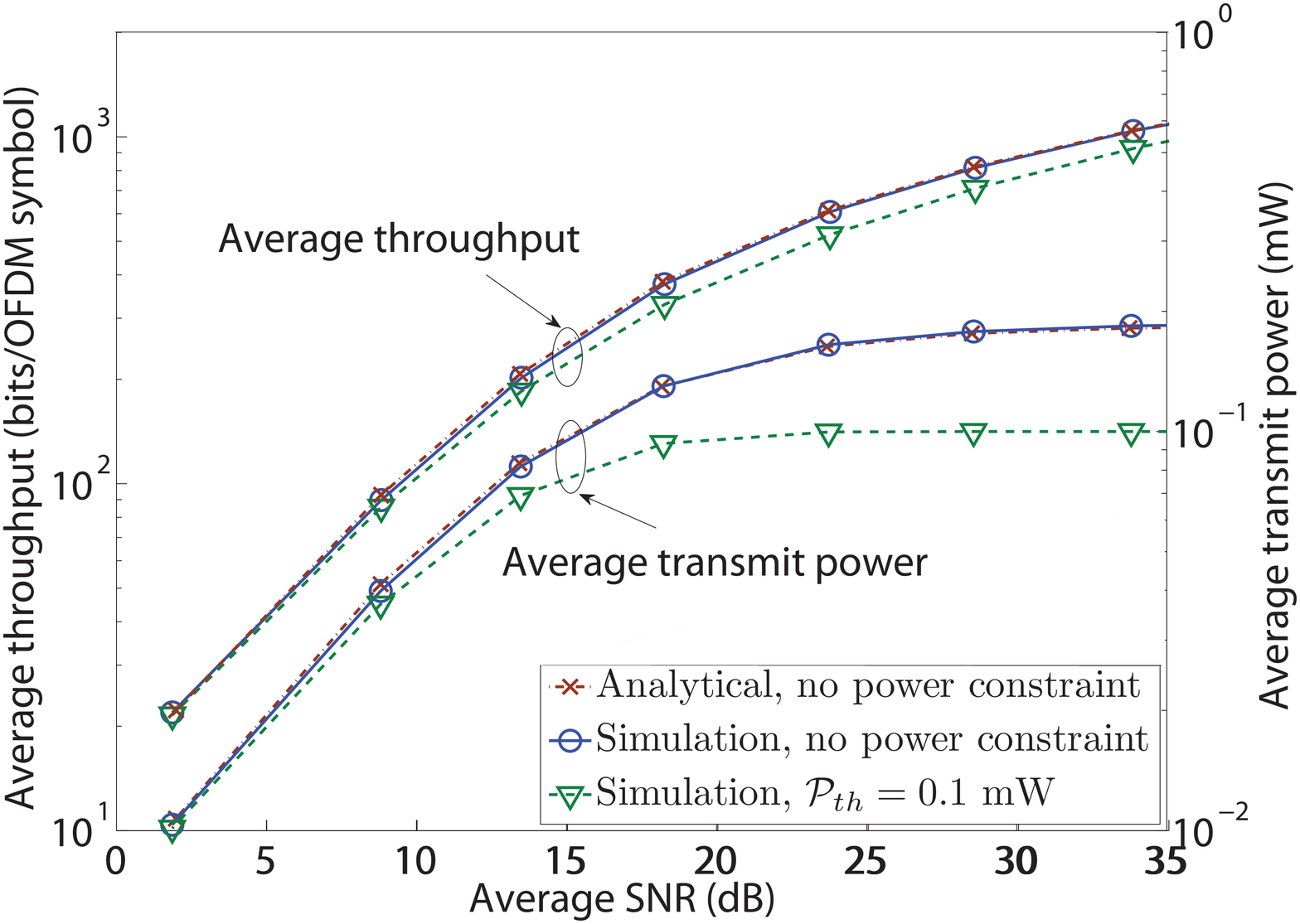}
	\caption{Average throughput and average transmit power as a function of average SNR, with and without a power constraint.}
	\label{fig:proposed}
\end{figure}

\begin{figure*}[t]
\begin{minipage}[b]{0.32\linewidth}
\centering
\includegraphics[width=1.10\textwidth]{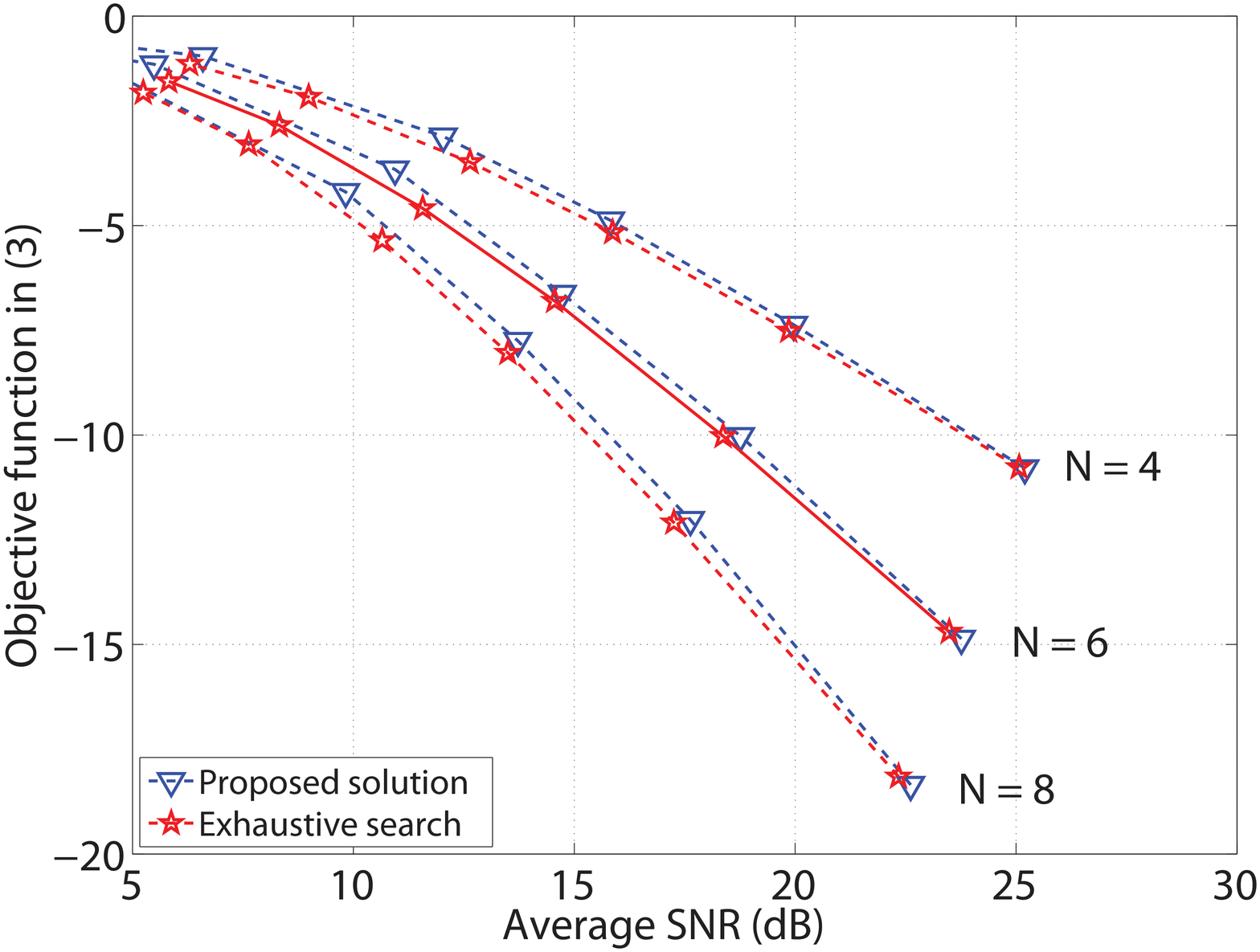}
\caption{Objective function for the proposed algorithm and the exhaustive search when $N$ = 4, 6, and 8.}
\label{fig:ex}
\end{minipage}
\hspace{0.3cm}
\begin{minipage}[b]{0.32\linewidth}
\centering
\includegraphics[width=1.10\textwidth]{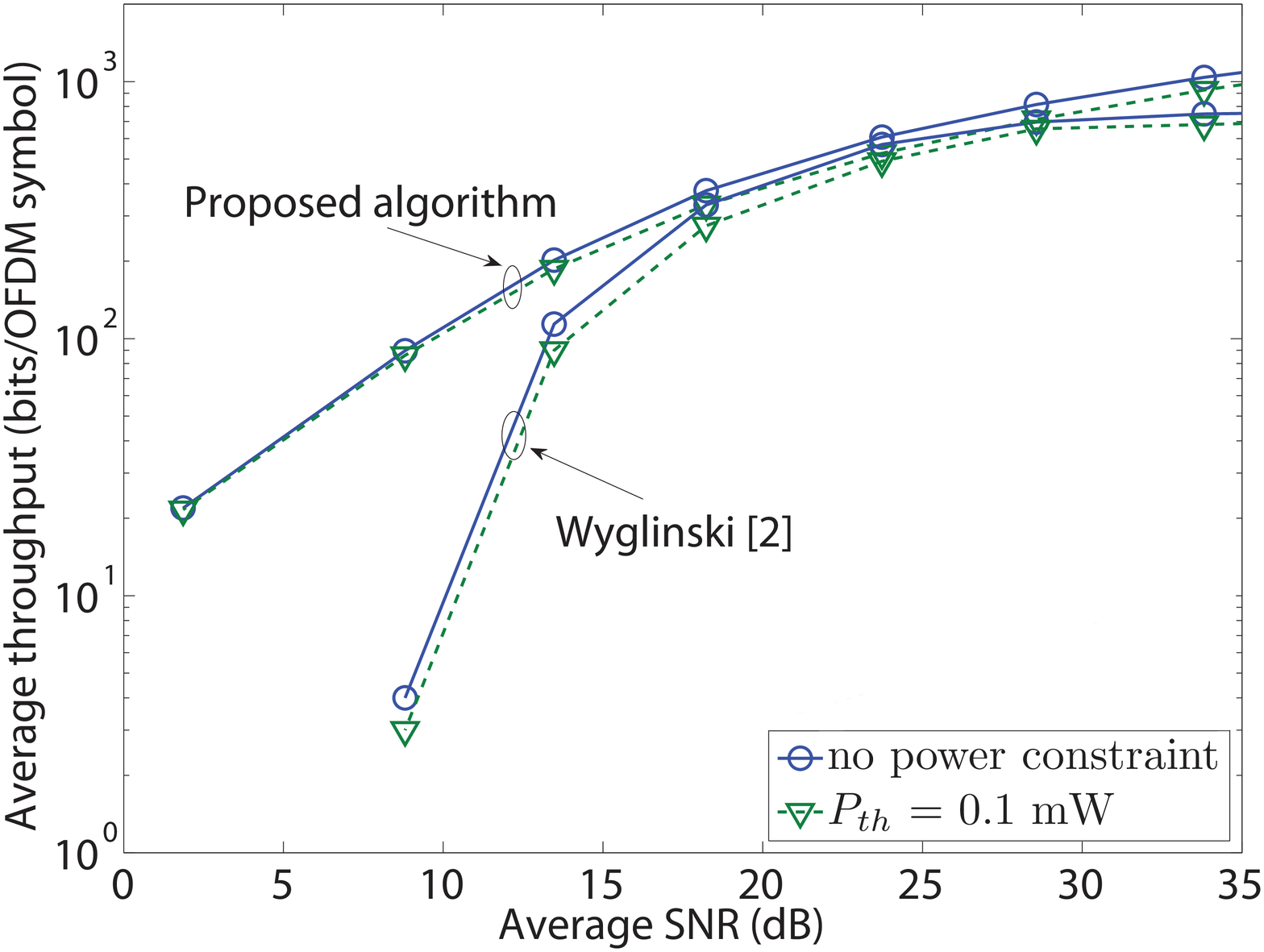}
\caption{Average throughput as a function of average SNR for the proposed algorithm and Wyglinski's algorithm in \cite{wyglinski2005bit}.}
\label{fig:Throughput_comp}
\end{minipage}
\hspace{0.3cm}
\begin{minipage}[b]{0.32\linewidth}
\centering
\includegraphics[width=1.10\textwidth]{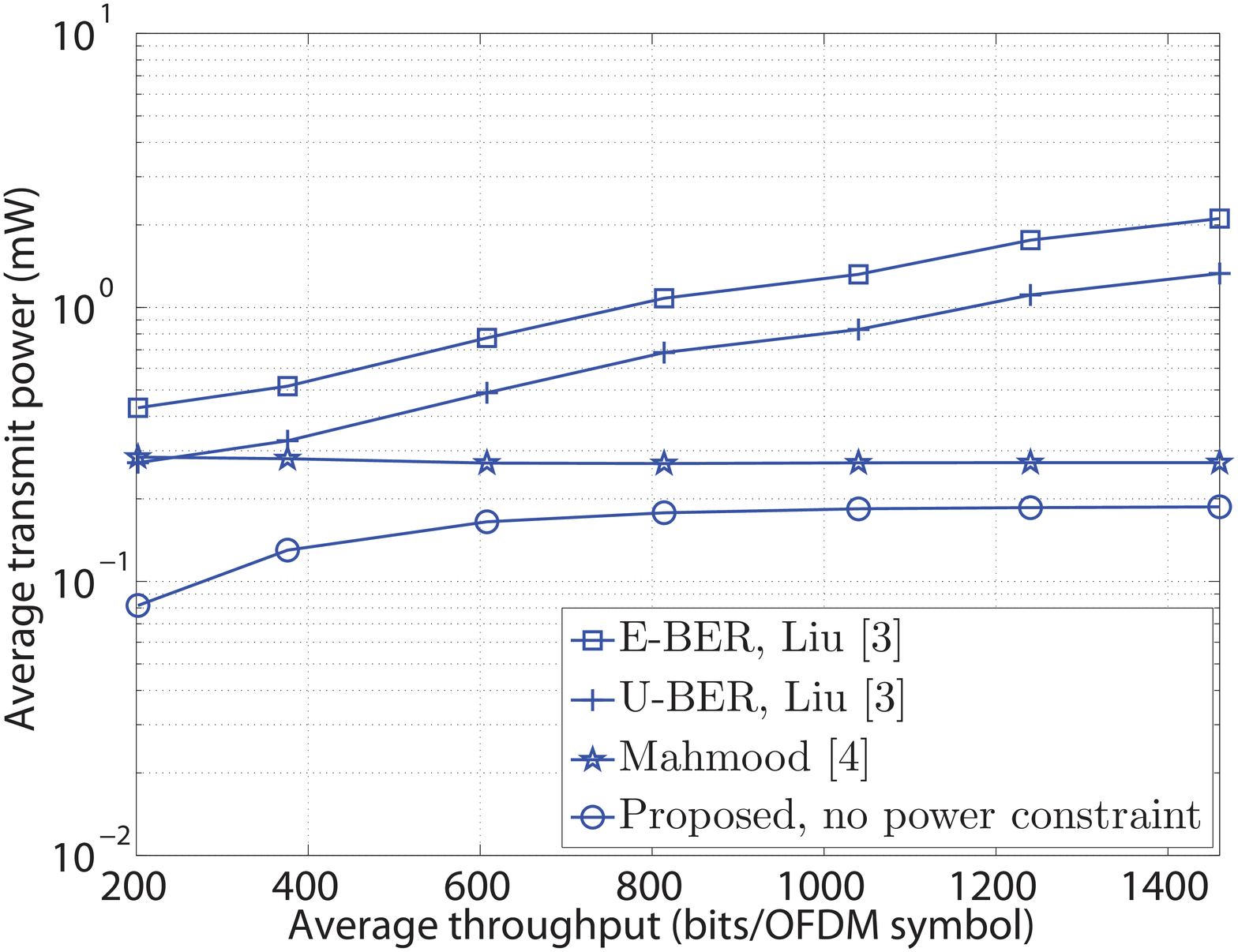}
\caption{Average transmit power as a function of average throughput for the proposed algorithm and the algorithms in \cite{liu2009adaptive} and \cite{mahmood2010efficient}.}
\label{fig:power}
\end{minipage}
\end{figure*}
Fig. \ref{fig:proposed} depicts the average throughput and transmit power as a function of the average SNR\footnote{The average SNR is calculated by averaging the instantaneous SNR values
per subcarrier over the total number of subcarriers and the total number of
channel realizations, respectively.}, with and without considering the total power constraint. In the latter case, the average throughput and transmit power, obtained by averaging (\ref{eq:eq10}) and  (\ref{eq:eq12}), respectively, over the total number of channel realizations through Monte Carlo simulations, show an excellent match to their counterparts in (\ref{eq:av_th}) and (\ref{eq:av_power}), respectively.  Further, for an average SNR $\leq$ 24 dB, one finds that both the average throughput and transmit power increase as the SNR increases, whereas for an average SNR $\geq$ 24 dB, the transmit power saturates while the throughput continues to increase. This observation can be explained as follows. The relation between $b_i$ and $\mathcal{P}_i$ in (\ref{eq:eq8}) implies that increasing the number of bits at the low range of $b_i$ (that exists at low average SNR values) occurs at the expense of additional transmit power, while increasing the number of bits at the high range of $b_i$ (that exists at high average SNR values) occurs at negligible increase in the transmit power. Accordingly, for lower values of the average SNR, increasing the average throughput is accompanied by a corresponding increase in the transmit power. On the other hand, for higher values of the average SNR, the average transmit power saturates and the average throughput is increased.
By considering a total power constraint, $\mathcal{P}_{th}$ = 0.1 mW, at lower SNRs, when the total transmit power is below the threshold, the average transmit power and throughput are similar to their respective values for the no power constraint case.
As the SNR increases, the transmit power reaches the power threshold and the average throughput is reduced accordingly.
Fig. \ref{fig:ex} compares the objective function achieved with the proposed algorithm and the exhaustive search that finds the discretized global optimal allocation for the problem in (\ref{eq:ineq_const}). Results are presented for $\mathcal{P}_{th}$ = 5 $\mu$W and $N$ = 4, 6, and 8; a small number of subcarriers is chosen, such that the exhaustive search is feasible. As can be seen, the proposed algorithm approaches the optimal results of the exhaustive search, and, hence, provides a close-to-optimal solution.

\vspace*{-5pt}
\subsection{Performance Comparison with Algorithms in the Literature}
In Fig. \ref{fig:Throughput_comp}, the throughput achieved by the proposed algorithm is compared to that obtained by Wyglinski's algorithm \cite{wyglinski2005bit} for the same operating conditions, with and without considering the total power constraint. For a fair comparison, the uniform power allocation used by the allocation scheme in \cite{wyglinski2005bit} is computed by dividing the average transmit power allocated by our algorithm by the total number of subcarriers. As shown in Fig.~\ref{fig:Throughput_comp}, the proposed algorithm provides a significantly higher throughput than the scheme in \cite{wyglinski2005bit} for low average SNRs. This result demonstrates that optimal allocation of transmit power is crucial for low power budgets.

Fig. \ref{fig:power} compares the average transmit power obtained by the proposed algorithm, in the case of no power constraint, with the optimum power allocation of Liu and Tang \cite{liu2009adaptive}, a variation called E-BER \cite{liu2009adaptive} that assumes an equal BER per subcarrier, and the algorithm of Mahmood and Belfiore \cite{mahmood2010efficient}. 
After matching the operating conditions, one can see that the proposed allocation scheme assigns less average power than the schemes in \cite{liu2009adaptive} and \cite{mahmood2010efficient} to achieve the same average BER and throughput. The different results between \cite{liu2009adaptive} and \cite{mahmood2010efficient} (while both guarantee the same fixed throughput) are mainly because the algorithms in \cite{liu2009adaptive} allocate the same number of bits per subcarrier, while the algorithm in \cite{mahmood2010efficient} allocates a different number of bits per subcarrier, which is intuitively more efficient.

The improved performance of the proposed joint bit and power allocation algorithm does not come at the cost of additional complexity. Its computational complexity is of $\mathcal{O(N)}$ when the initial value of $\alpha$ results in an inactive power constraint, which is similar to that of Liu's algorithm. Otherwise, it is of  $\mathcal{O}(\mathcal{N} \textup{log} (\mathcal{N}))$, which is lower than that of Wyglinski's $\mathcal{O}(\mathcal{N}^2)$ algorithm and significantly lower than $\mathcal{O}(\mathcal{N}!)$ of the exhaustive search.


\section{Conclusion} \label{sec:conc}
In this letter, we proposed a novel algorithm that jointly maximizes the throughput and minimizes the transmit power given constraints on the BER per subcarrier and the total transmit power. Closed-form expressions were derived for the close-to-optimal bit and power allocations per subcarrier, average throughput, and average transmit power. Simulation results demonstrated that the proposed algorithm outperforms different allocation schemes that separately maximizes the throughput or minimizes the transmit power, under the same operating conditions, while requiring similar or reduced computational effort. Additionally, it was shown that its performance approaches that of the exhaustive search with significantly lower complexity.
\vspace*{-6.5pt} 

%

\ifCLASSOPTIONcaptionsoff
  \newpage
\fi

\end{document}